\documentclass{article}
\usepackage{amssymb}
\usepackage{epsfig}

\newcommand{\be}{\begin{equation}}
\newcommand{\ee}{\end{equation}}

\begin{document}

\begin{flushright}
Liverpool Preprint: LTH 442\\
 \end{flushright}
  
\vspace{5mm}
\begin{center}
{\LARGE\bf Quark mass dependence of hadron masses from lattice QCD}\\[10mm] 
{\large\it UKQCD Collaboration}\\[3mm]
 
{\bf M. Foster and  C. Michael}\\

{Theoretical Physics Division, Department of Mathematical Sciences, 
University of Liverpool, Liverpool, L69 3BX, U.K.}\\[2mm]

\end{center}

\begin{abstract}

 We discuss lattice methods to obtain the derivatives of a lattice meson
mass with respect to the bare sea and valence quark masses. Applications
 are made to quenched and dynamical fermion configurations. We find 
evidence for significant differences between quenched and dynamical
fermion configurations. We discuss how to relate dependence on the bare
lattice parameters  to more phenomenologically useful quantities.

\end{abstract}

\section{Introduction}

  In lattice studies of QCD, the action depends on several bare
parameters  such as the inverse coupling $\beta$ and those controlling
the quark masses. Here we distinguish  the sea quarks which contribute
to the vacuum and valence quarks which propagate in the  vacuum but do
not contribute to it. Thus there will be two possible parameters 
describing the quarks masses: the sea and valence hopping parameters
($\kappa_s$  and $\kappa_v$). In lattice studies, unlike experiment, it
is  possible to vary each of these mass parameters independently.

It is of interest to  establish the dependence of quantities of physical
interest, such as  hadron masses, on these quark mass parameters. For
instance, the valence-quark mass dependence  of the meson mass controls
the $J$ parameter which is related~\cite{J} to the slope  of $M_V$
versus $M_P^2$ (where $M_P$ and $M_V$ are the pseudoscalar and vector
meson masses respectively). This slope is  found in lattice studies to
be significantly  smaller than the experimental value. It is a challenge
for dynamical fermion  studies on a lattice to narrow this discrepancy
as the sea quark mass is reduced.
 Another area of current interest is the magnitude of  sea quark effects
 on hadron masses. The dominant effect of sea quarks is just to 
renormalise the coupling ($\beta$) so it is valuable to have techniques 
to explore in fine detail the sea quark effects so that physically 
significant effects can be explored in  dynamical fermion studies.

One direct way to achieve this is to study the theory at many different
combinations of parameters. This is the conventional way  to  study the
valence quark mass dependence and is reasonably efficient since the 
lattice configurations themselves do not depend on $\kappa_v$. For the
sea quark  mass, however, this is a computationally challenging
endeavour since  different gauge configurations must be constructed for
each $\kappa_s$ value  and then the finite differences of hadron masses
between these different ensembles  of configurations will be small and
quite noisy. 

One way to obtain estimates of  derivatives by working with a lattice 
ensemble at one set of parameters is described in ref\cite{match}. Here 
we specialise to  explore a method to obtain the derivative of a hadron
mass with respect to a parameter such as $\kappa_s$.  The method is
essentially to  take formally the derivative of a lattice identity. This
method, often called a `sum rule',  has been used before to obtain
derivatives with respect to $\beta$~\cite{cmsumrule}. Here we use a
similar approach to extract derivatives with respect to $\kappa_s$ and
$\kappa_v$ - see also \cite{wuppk}.

 The derivative with respect to $\kappa_v$ involves a three point
function of  fermion fields and so cannot be obtained from propagators
from one source only. Here we choose to use stochastic
propagators~\cite{roma-tv96} with maximal variance
reduction~\cite{cmpeisa}  which allow the appropriate propagator
combination to be evaluated. 

 For the derivative with respect to $\kappa_s$, a disconnected three
point  function is needed. In this case we use Z2 noise
methods~\cite{z2liu,sesam} to evaluate  the appropriate combination of
propagators. We apply this to quenched  and dynamical fermion gauge
configurations and see a significant difference. We discuss the impact of
these results on the sea-quark dependence of  meson masses.

 This study is exploratory and we discuss the computational effort needed 
to extract these derivatives with respect to bare quark masses. We also 
compare our results with those obtained by taking finite differences.

\section{Quark mass dependences}

 The mesonic masses in lattice studies are determined by measuring
two-point  correlations of appropriate operators at large time
separation $t$. We then  wish to take the formal derivative with respect
to a parameter representing  the quark mass. This will give the required
sum rules for the derivative  of the lattice hadron mass with respect to
the quark mass parameter.

Consider an action density
 \be
  S=S_f + \beta S_g = \sum_1^{N_f} \bar{\psi} {\cal M} \psi + \beta S_g
 \ee
 where, for the Wilson-Dirac discretisation of fermions, 
 \be
   {\cal M}=m + D
 \ee
 where the quark mass parameter  $m \equiv 1/\kappa$,  with $\kappa$ 
the conventional hopping parameter,  so  in terms of the bare quark mass
$m_b$ in the naive continuum limit $m=8+2am_b$. The  term $D$ contains
the Wilson nearest neighbour  gauge link terms as well as the SW-clover
terms with coefficient $C_{SW}$.
  The hadronic correlation is then given by
 \be
   C(t) = {1 \over Z}  \int H(0) H^{\dag}(t) e^{S}
 \ee
 Here for mesons $H$ will be of the form $\bar{\psi} \Gamma \psi$ and 
in this work we will concentrate on the case of flavour non-singlet
mesons  so that the hairpin diagrams will not be needed. Then  the
fermionic degrees of freedom are integrated out giving a factor  of the
inverse of the fermion matrix ($G={\cal M}^{-1}$) for each pairing:
 \be
    C(t) = \langle 0| \left( G(0,t) \Gamma G(t,0) \Gamma \right) |0 \rangle
 \ee   
 At large $t$, this correlation will be dominated by the ground state
meson  with the quantum numbers created by $H$:
 \be
   C(t)=c_0^2 e^{-M_0 t} + \dots
 \ee
 This sketch of the formalism allows us to explore taking the derivative
 with respect to the  quark mass parameter $m$ (actually the inverse
hopping parameter) on each side of the above expressions for $C(t)$.
This derivative  is to be taken at fixed $\beta$. Then, since formally
the only $m$-dependence is in  the exponent, the derivative brings down
a factor of $N_f \bar{\psi} \psi$.
 \be
  {dC(t) \over dm} ={1 \over Z}  \int H(0) H^{\dag}(t) N_f \bar{\psi}
\psi e^{S}    - C(t) {1 \over Z}  \int N_f \bar{\psi} \psi e^{S}
 \ee
 where the second term comes from the $m$-dependence implicit in $Z$. On
integrating  out the six fermions, this  will give two diagrams,
connected and disconnected (actually only the connected part of the
disconnected diagram  will contribute as discussed below). Thus, summing
explicitly over the  insertion at $t_1$, we have 
 \be
  {dC(t) \over dm} =\sum_{t_1} \left(-C_3(t_1,t)+N_f D_3(t_1,t) \right)
 \ee

\begin{figure}[htb]
\begin{center}
\epsfig{figure=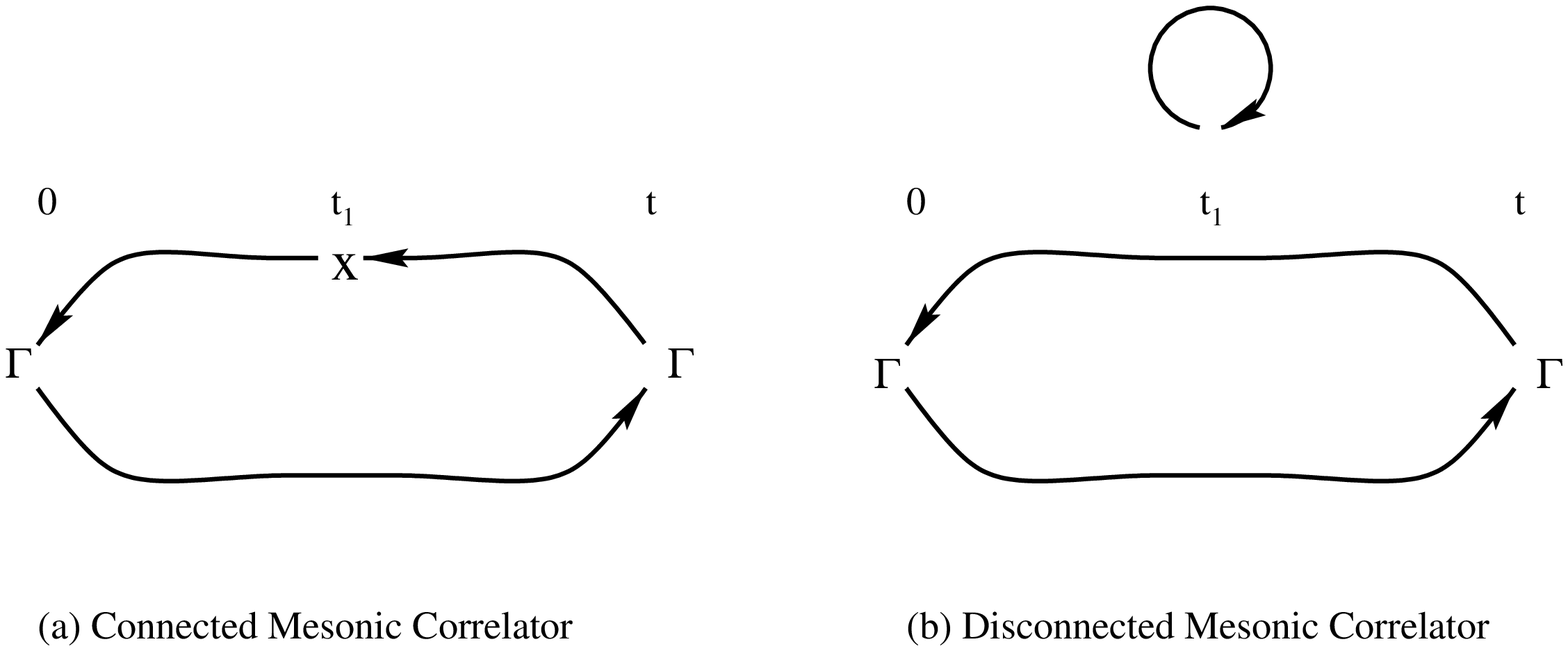,height=2.0in}
\end{center}
\caption{}
\label{disc.fig} The diagrams corresponding to $C_3$ and $D_3$.
\end{figure}

 Where for the connected diagram there will be  terms from the insertion
on either quark line:
 \be
 C_3(t_1,t)=\langle 0|  \left( G(0,t_1)G(t_1,t) \Gamma G(t,0) \Gamma
\right)  +  \left( G(0,t) \Gamma G(t,t_1) G(t_1,0) \Gamma
\right) | 0 \rangle
 \ee
  while the disconnected diagram is, for each flavour of quark in the loop,   
 \be
    D_3(t_1,t) =  \langle 0| \left( G(0,t) \Gamma G(t,0) \Gamma \right) 
\left( G(t_1,t_1) \right) |0 \rangle -  \langle 0| \left( G(0,t) \Gamma
G(t,0) \Gamma \right) |0  \rangle 
 \langle 0| \left( G(t_1,t_1) \right) |0\rangle 
 \ee
 where curved brackets imply a trace over colour, spin and space coordinates.

In the quenched approximation, only the connected diagram contributes.
This can be  seen another way since ${\cal M}G=1$ implies ${\cal
M} dG/dm+ d{\cal M}/dm \, G =0$ and since  $d{\cal M}/dm=1$ by definition,
then $dG/dm=-{\cal M}^{-1}G=-GG$. Thus either fermion  propagator in the
mesonic correlator can be `opened' by an insertion.

For dynamical fermions, both types of diagram contribute but one can see
that the  connected diagram corresponds to varying the valence quark
mass while the  disconnected diagram corresponds to varying the sea
quark mass.

The two point hadronic correlation can be expressed in terms of a sum over 
intermediate  states of masses $M_i$
 \be
 C(t)=\sum_{i} c_i^2 e^{-M_i t}
 \ee
The leading term in the derivative at large $t$ can be then be evaluated
 \be
  {dC(t) \over dm} =  -t {dM_0 \over dm} c_0^2 \, e^{-M_0 t}+
{dc_0^2 \over dm} e^{-M_0 t} + \dots
 \ee
 and it thus behaves as  $t e^{-M_0 t}$ where $M_0$ is the ground state
meson mass.

 We now extract the contribution from the right hand side which has this
same behaviour. For both $C_3$ and $D_3$,  the insertion is summed over
all space and time.  Then the leading term   arises when the lightest
allowed meson propagates and when  $0 < t_1 < t$. This will produce
terms which are linear in $t$ which  arise from the $t$ possible 
insertions (at $t_1$) between the  creation and destruction of the
meson. Then evaluating this  ground state meson contribution,  for the
connected diagram, gives
 \be
 \sum_{t_1} C_3(t_1,t) = c_0^2 \, t \left( X^{(1)}+X^{(2)} \right) 
e^{-M_0 t}=t\, \left( X^{(1)}+X^{(2)}\right) C(t)
 \ee
 where the suffix refers to the insertion on quark propagator 1 or 2 and
$X$  is the matrix element of the $\bar{\psi}\psi$ insertion between
ground state hadrons.  A similar expression applies for the disconnected
case.

Equating the coefficients of the  terms behaving as $t e^{-M_0 t}$ on each 
side of the identity, then gives the exact result that
 \be
   {dM_0 \over dm_v} =  X^{(1)}+X^{(2)}
 \ee
 where the matrix element sum can be obtained by extracting the ground
state contribution to $C_3/C$. In principle this can be obtained by
taking  the insertion in $C_3$  such that $0 < t_1 < t$ and  both $t_1$
and $t-t_1$ are large so that the ground state contributes. So 
we can write
 \be
   {dM_0 \over dm_v} =  \lim_{t_1,(t-t_1) \to \infty} {C_3(t_1,t) \over C(t)}
 \ee

 This sum rule  relates the derivative to an expression that can be
evaluated  from lattice configurations at only one set of parameters. It
is an  exact  identity. If there is a dependence of  the lattice meson
mass $M_0$   on the finite spatial size $L$ of the lattice, the
derivative should be taken at fixed number of lattice spacings, not at
fixed physical size. 
 These considerations are very similar to those used in the
lattice sum rules derived  by taking formal derivatives with respect to
$\beta$~\cite{cmsumrule}.

 For the disconnected diagram, the equivalent expression is 
 \be
   {dM_0 \over dm_s} = - N_f \lim_{t_1,(t-t_1) \to \infty}
    {D_3(t_1,t) \over C(t)}
 \ee

 In order to evaluate these expressions on a lattice, it is useful  to
consider efficient  ways in which excited state contributions can be
eliminated,  since the formal limits of large $t$ will have big noise to
signal. Here we consider the connected correlation $C_3$ and use a
complete set of hadron states  of mass $M_i$  in the intermediate
intervals of time extent $t_1$ and $t_2=t-t_1$.

In practice, we will be using more than one operator to create  and
destroy the hadronic state. This allows an optimal combination of  these
operators to be formed that minimises the excited state contribution.
Then the two-body correlation between  operators $a$ at $t=0$ and $b$ at
$t$ will be given by 
 \be
   C^{(ab)}(t)=\sum_{i} c^{(a)}_i e^{-M_i t} c^{(b)}_i
 \ee
 \be
   C_3^{(ab)}(t_1,t_2)=\sum_{i,j} c^{(a)}_i e^{-M_i t_1} x_{ij} 
                  e^{-M_j t_2} c^{(b)}_j
 \ee
  where $x_{00}$ is the required quantity ($X^{(1)}+X^{(2)})$ - the
matrix element appropriate  to  the ground state meson of mass $M_0$. 
We might expect $x_{11}$  to be similar in sign and magnitude  to
$x_{00}$ if the quark mass dependence of the  excited state is
comparable to that of the ground state and thus excited state
contributions would  cancel in the ratio $C_3/C$. This is incorrect,
since the off-diagonal  terms ($x_{01}$) will dominate the excited state
contributions to $C_3$ since the excited state only  propagates for the
shorter interval $t_1$ (or $t_2$). One way to extract  $x_{00}$ is to
make a fit to the three point data with both $t_1>t_{\rm min}$ and
$t_2>t_{\rm min}$, keeping the coefficients  and masses ($c^{(a)}_i$ and
$M_i$) fixed from the fit to the two-point function  data with $t>t_{\rm
min}$. This can be compared with the more direct approach of looking 
for a plateau in $C_3(t_1,t_2)/C(t)$ as $t_1$ and $t_2$ are  increased
(with $t=t_1+t_2$). 

 When  two (or more) different types of hadronic creation operators are
used, a variational method is  an effective way to determine the ground
state contribution to $C$ and  hence to extract the ground state
contribution to  $C_3$.  Alternatively, if a two state fit to the
two-body correlation between two operators  at each end is made, then 
from the coefficients, it follows that a combination of operators 
$c^{(2)}_1 H_1 - c^{(1)}_1 H_2$ will remove the contribution of the
excited state  in the approximation that only two states  contribute to
the correlations. Then
 we can use this combination to evaluate the ground state component of
$C_3/C$, using 
 \be
 {c_1^{(2)} c_1^{(2)} C_3^{(11)} -2 c_1^{(2)} c_1^{(1)} C_3^{(12)} +
      c_1^{(1)} c_1^{(1)} C_3^{(22)}
 \over c_1^{(2)} c_1^{(2)} C^{(11)} -2 c_1^{(2)} c_1^{(1)} C^{(12)} +
      c_1^{(1)} c_1^{(1)} C^{(22)}}
  \ee
 A similar analysis holds equivalently for the extraction  of the ground
state contribution $d_{00}$ to $D_3$.

 When $t \approx T/2$, the contributions from propagation around the
time boundary  of the lattice may be significant. For $C_3$ there will
be no  such `round the back' term because the insertion is made
explicitly, while for $D_3$  the  connected matrix element involved 
will cancel  for the round the back term.   In contrast the two-body
correlator $C$ will be a sum of two terms. Illustrating this for  the
ground state component for one type of operator, we have:
 \be
 C= c_0^2 e^{-M_0 t} +c_0^2 e^{-M_0 (T-t)}
\ee
 \be
 C_3= c_0^2 e^{-M_0 t} x_{00}
\ee
 Hence
 \be
 C_3/C= x_{00}{1 \over 1+e^{M_0(T-2t)} }
\ee
 This formalism can be used to correct for the different $t$-dependences
when  looking for a plateau in $C_3/C$ and in $D_3/C$ as $t$ increases.

 We now discuss efficient methods to evaluate these correlators on a
lattice.

\section{Valence quark mass dependence}

As a first application, we consider the dependence of the hadron mass 
on the valence quarks. For this a three point function needs to be
evaluated - see fig.~1a.  Thus conventional quark propagators from one
source are inadequate  for this task. One feasible way forward is to use
a stochastic inversion method  which allows the evaluation of quark
propagators from any site to any other site. Although the stochastic
method  is not more efficient than the conventional inversion from one
source  for mesons made of light quarks~\cite{cmpeisa}, it does allow
the flexibility  to evaluate three point correlations readily. For this
reason it allows  an exploratory study of this area.

Stochastic propagators~\cite{cmpeisa,roma-tv96} are one technique to
invert the fermionic matrix for the light quarks. They can be used in
place of light quark propagators calculated with the usual deterministic
algorithm.  The stochastic inversion is based on the relation:
 \begin{equation}
        G_{ij} =  {\cal M}_{ij}^{-1}=\frac1Z \int {\cal D}\phi\;
        ({\cal M}_{jk}\phi_k)^\ast \phi_i\; 
        \exp \left( -\phi_i^\ast ({\cal M}^\dagger {\cal M})_{ij}
        \phi_j \right) 
 \end{equation}  where, in our case, ${\cal M}$ is the improved
Wilson-Dirac fermionic operator and the indices $i,j,k$ represent
simultaneously the space-time coordinates, the spinor and colour
indices.  For every gauge configuration, an ensemble of independent fields
$\phi_i$ (we use 24 following~\cite{cmpeisa}) is generated with gaussian
probability:
 \begin{equation} 
P[\phi] =\frac1Z \exp \left(
-\phi_i^\ast ({\cal M}^\dagger {\cal M})_{ij} \phi_j \right) 
 \end{equation} 
All light propagators are computed as averages over the
pseudo-fermionic samples:
\begin{equation}
        G_{ij} = 
        \left\{\begin{array}{l}
               \langle ({\cal M}\phi)_j^\ast \phi_i \rangle \\
               \mathrm{or} \\
               \gamma_5 \langle \phi_j^\ast ({\cal M}\phi)_i \rangle\gamma_5 
               \end{array}
        \right. \label{eq:stock} 
 \end{equation} 
 where the two expressions are related by $G_{ij} = \gamma_5
G_{ji}^\dagger \gamma_5$.  Moreover, the maximal variance reduction
method is applied in order to minimise the statistical
noise~\cite{cmpeisa}. The maximal variance reduction method involves
dividing the lattice  into two boxes ($0<t<T/2$ and $T/2<t<T$) and
solving the equation of motion numerically within each box, keeping the
pseudo-fermion field $\phi$ on the boundary fixed.  According to the
maximal reduction method, the fields which enter the correlation
functions must be either the original fields $\phi$ or solutions of the
equation of motion in disconnected regions.  The stochastic propagator
is therefore defined from each point in one box to every point in the
other box or on the boundary. For this reason, when computing the
three-point correlation function, 
 \begin{equation}
  \sum_{x,y,z} \langle 0 | H(t_1,x) {\cal O}(t_0, y) 
          H^\dagger(t_2,z) | 0 \rangle 
 \end{equation}  
 the  operator   ${\mathcal O}$ (which  is $\bar{\psi} \psi$) is forced
to be on the boundary ($t_0=0$ or $T/2$) and the other two operators
must be in different boxes, while the spatial coordinates are not
constrained. If $j$ is a point of the boundary, not all the terms in
$({\cal M}\phi)_j$ lie on the boundary because the operator ${\cal M}$
involves first neighbours in all directions. Hence, whenever a
propagator $G_{ij}$ is needed with one of the points on the boundary, we
use whichever of the two expressions in Eq.~\ref{eq:stock} has 
${\cal M}\phi$ computed away from the boundary. This implies that 
we are restricted to $t \ge 2$.

The numerical analysis used 24 stochastic samples  on each of 20
quenched gauge configurations, generated~\cite{cmpeisa} on a $12^3
\times 24$ lattice at $\beta=5.7$, corresponding to $a^{-1}=0.91$ GeV. 
With improved clover coefficient $C_{SW}=1.57$, we use two values of
$\kappa$: $\kappa_1=0.14077$ and $\kappa_2=0.13843$. The lighter value
$\kappa_1$ corresponds to a bare mass of the light quark around the
strange mass. The chiral limit corresponds to
$\kappa_c=0.14351$~\cite{hugh97}. Error estimates come from bootstrap 
over the gauge configurations.  We also made an exploratory study  of
some dynamical fermion configurations, as will be discussed later.

In smearing the hadronic interpolating operators, spatial fuzzed links
are used. Following the prescription in~\cite{cmpeisa, michael95},
to which the interested reader should refer for details, the fuzzed
links are defined iteratively as:
\begin{equation}
        U_{\mathrm{new}} = {\mathcal P} \left(f U_{\mathrm{old}} 
          + \sum_{i=1}^4 U_{\mathrm{bend},i} \right) 
 \end{equation}
 where ${\mathcal P}$ is a projector over $SU(3)$, and
$U_{\mathrm{bend},i}$ are the staples attached to the link in the
spatial directions. Five iterations of fuzzing with $f=2.5$ are used and
then  the fuzzed links are combined to straight paths of  length three.
The fuzzed fermionic fields are defined following~\cite{michael95}.

\begin{figure}[bt!] 
\vspace{11cm} 
\includegraphics{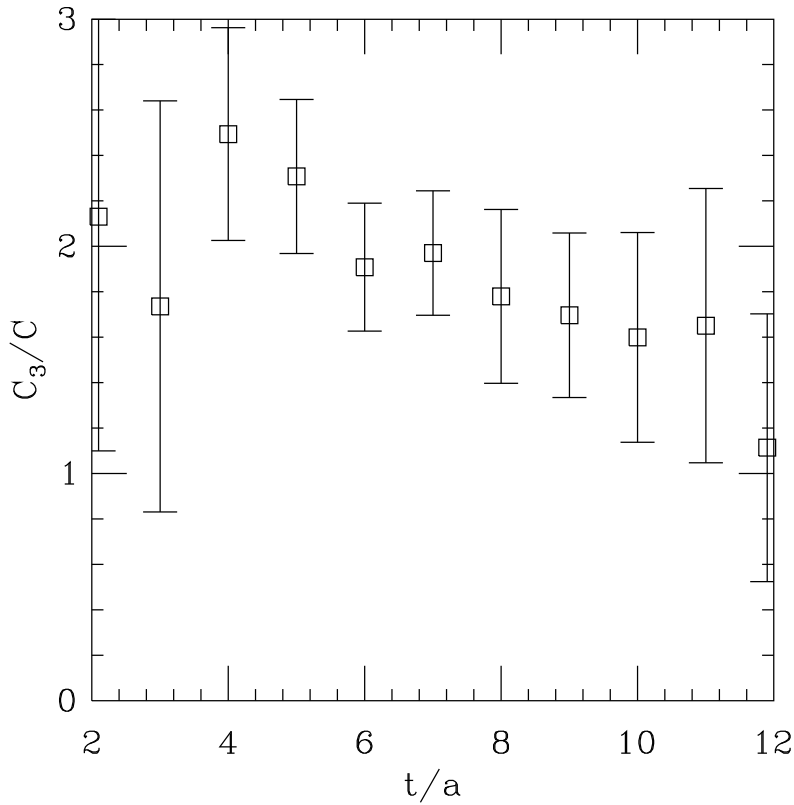}
 \caption{ The connected correlation $C_3/C$  versus
 time $t=t_1+t_2$ in lattice units. The  data are for the variational
combination that  reduces the excited state contribution and are from
quenched  lattices with $\kappa=0.14077$ with $|t_1-t_2| < 2$. We expect
the ground state contribution to be dominant when  $t_1 > 2$ and $t_2 >
2$, that is for $t \ge 6$.
   }
\label{fig.val}
\end{figure}

 We employed  two types of hadronic operator for the correlations -
local and fuzzed - yielding  a $2 \times 2$ matrix. From this we use a
variational approach to extract  the linear combination of operators
which maximises the ground state contribution  -  as described above.
Since we are able to get good  two state fits to the two-body
correlations for the pseudoscalar meson for $t \ge 3$, this  variational
linear combination was determined using $t$-values 3 and 4.
 In order to maximise the  ground state contribution relative to excited
states, we evaluated the three point  diagram $C_3(t_1,t_2)$ using 
values of $t_1$ and $t_2$ near to  $t/2$ where $t=t_1+t_2$. The ground
state improved ratio of $C_3(t_1,t_2/C(t)$ is plotted  in
fig.~\ref{fig.val}. The extraction of the ground state should be good  if
$t_1,t_2 \ge 3$. For odd values of $t$ there are higher statistics (from
the  3,4 and 4,3 partitions of t=7 for instance). Thus we expect $t=7$
to be the  best determined value and this is given  in
table~\ref{table.val}. Consistency at higher $t$-values confirms that
the  ground state extraction is correct.

\begin{table}[hbt]

\caption{Connected loop correlations.}

\begin{center}
\begin{tabular}{lllllc}
  $\kappa_v$ &  $ M_P$ &  $dM_P/dm_v$ &  $ M_V$ &  $ dM_V/dm_v$& 
  $n_{\rm gauge}$ \\
\hline
 0.14077 & 0.529(2)  & 1.97(27) &0.815(5) &  1.9(9)&20 \\ 
 0.13843 &0.736(2) &  1.56(22) & 0.938(3) &   0.3(4)&20 \\ 
\hline  
 0.1395  &0.558(8) &  1.3(3)  & 0.786(9) &   0.2(1.1)&5\\ 

\end{tabular}
\label{table.val}
\end{center}
\end{table}

 For the pseudoscalar meson, we expect that $M_P^2$ is approximately 
linear in $m_v$. Thus $dM_p/dm_v$ should decrease like $1/M_P$ which  is
indeed consistent with the results shown in table~\ref{table.val}.  From
the  high statistics spectroscopy at these two hopping
parameters~\cite{hugh97},  one can evaluate the finite difference
obtaining $dM_P^2/dm_v=2.18(4)$  which agrees very well with the values
determined from the sum rules of $2 M_P dM_P/dm_v= 2.08(29)$ and
2.30(32) at $\kappa_v=0.14077$ and 0.13843  respectively.

For the vector meson, the expectation is that $M_V$ is  approximately
linear in $m_v$.  The finite difference~\cite{hugh97} gives 
$dM_V/dm_v=1.02(7)$.  The sum rule determination with our current
statistics  is too noisy at $t=7$ to give an accurate value. For the 
heavier quark mass ($\kappa_v=0.13843$) a good two-state fit to the two
point correlation data can be made for  $t \ge 2$. This allows us to use
$t=4$ and 5 for the $C_3/C$ ratio and we obtain 0.90(18) and 0.93(26)
respectively, in excellent agreement with the  expected value. At the
lighter  quark mass, we need $t \ge 3$ for a two-state fit so the poor
result remains. This is a disappointment, since  from the values of
$dM_V/dm_v$ and $dM_P/dm_v$, one can evaluate the  $J$ parameter (which
is the physical quantity, defined in the continuum limit as $M_V
dM_V/dM_P^2$ at $M_V/M_P=1.8$) at a  quark mass corresponding to $m_v$.
Thus $J$ can be determined at the lightest quark mass directly, rather
than as a difference between two quark masses. This $J$ parameter is a
useful indicator~\cite{J} of the distance between quenched  QCD (with
$J\approx 0.37$) and experiment (with $J =0.48(2))$. Hence a quick and
accurate method to determine $J$ would be useful to  calibrate dynamical
fermion studies.

 We also evaluated the same quantities for  dynamical fermion
configurations~\cite{ukqcddf} at $\beta=5.2$ with two flavours of sea
quarks at $\kappa_s=0.1395$ on a  $12^3 24 $ lattice using a SW-clover 
improved action with $C_{SW}=1.76$. The correlation was evaluated with
$\kappa_v=\kappa_s$.  In this case the higher statistics determination
of the masses~\cite{ukqcddf} allows the derivative at fixed  $\kappa_s$
to be evaluated, giving  $dM_P/dm_v=2.2(4)$ from the finite difference
between $\kappa_v$ of 0.1395 and 0.1390. Our analysis is from only 5
gauge configurations and so the  error  may   be underestimated because
of the small sample size. For the pseudoscalar, we find acceptable two
state fits for $t \ge 2$ (the hadronic operators are local and fuzzed
with straight paths of 2 links) and so we may use the variational method
from $t$ of 2 to 3  to determine the ground state couplings. This 
yields a value  at t=4 and 5 of $C_3/C$ of 1.4(4) and 1.5(3),
respectively.  This is consistent   with the value in 
table~\ref{table.val} and with the finite difference value within
errors.  
 The vector meson case is too noisy to be of any use.  The main
conclusion is that the  ratio of correlations  $C_3/C$ is very similar
in the dynamical configurations to the quenched case. This is not really
surprising since the sea quark masses used in the dynamical quark study 
are fairly large -  larger than the strange quark mass.

\section{Sea quark mass dependence}

 The disconnected diagram (see fig.~1b) involves measuring two gauge
invariant  contributions: the two-point hadronic correlator $C$ and the 
loop contribution corresponding to ${\rm Trace}{\cal M}^{-1}$ where the 
trace is a sum over colour, spin and spatial coordinates at a given 
time $t_0$. This needs the propagator from each site on a time slice to 
a sink corresponding to the same site. There is an efficient way to
evaluate this  making use of Z2 stochastic sources~\cite{z2liu,sesam}.
Here we propose a variant of this method  which is appropriate for our
current study. This method also gives the two  point correlator
$C(t_1,t_2)$ for pseudoscalar mesons and vector mesons from any time
$t_1$  to any other time $t_2$. Then combined with the loop contribution
at $t_0$,  we have the ingredients needed to evaluate the required
connected part $D_3$ of the disconnected correlation.

 Details of the Z2 method used are given in the Appendix. In this
exploratory study on $12^3 24$ lattices,  we use local operators to
create the pseudoscalar and vector mesons. We have used rather generous
values of the number  of Z2 samples per time slice (namely between 16
and 32 for each of the  two related types of source used). This amounts
to 768 or more inversions (equivalent to  64 conventional propagator
inversions from 12 colour spin sources) per  gauge configuration.
Because of the decreased number of iterations  of the inversion
algorithm in our case, the time used is equivalent to about 30
conventional propagator determinations  per gauge configuration. This is
a substantial computational challenge, but  it does provide a
significant resource: the loop contributions at each $t$  and the
pseudoscalar and vector correlators from any $t_1$ to any $t_2$. Because
 of our choice of number of Z2 samples, we have negligible errors 
coming from the Z2 noise for the value of  ${\rm Trace}{\cal M}^{-1}$
from each time-slice and for  the pseudoscalar correlator from $t_1$ to
$t_2$. For the vector meson correlator, the error  from the Z2 method is
in some cases comparable to the intrinsic variation and we  correct for
this in derived quantities by increasing our errors appropriately where
necessary. Indeed, in retrospect, it would  have been more efficient for
the present study to use less Z2 samples and to explore more  gauge
configurations. Our approach, however, was that so much computational
effort  has gone into the production of the dynamical fermion
configurations that  the large number of inversions used in measurement
are in effect a  relatively small extra overhead.

 We  evaluated these  quantities for  dynamical fermion
configurations~\cite{ukqcddf} at $\beta=5.2$ with two flavours of sea
quarks at $\kappa_s=0.1390,\ 0.1395$ and 0.1398 on a  $12^3 24 $ lattice
using a SW-clover  improved action with $C_{SW}=1.76$. In our 
evaluations  we restrict ourselves to the case  where the propagating
quarks have the sea-quark mass, i.e.  $\kappa_v=\kappa_s$. The number
of gauge  configurations used and number of Z2 samples $n_Z$ are given in
table~\ref{table.disc}. We also quote, for completeness, the
pseudoscalar and vector meson masses and the $R_0$ values obtained from
higher statistics  by conventional methods~\cite{hugh97,r0,ukqcddf}.

 It is  possible to measure the disconnected diagram in  quenched  gauge
configurations as well as in dynamical fermion configurations. We use
the same gauge configurations as  discussed in the previous section. For
the quenched case, we include a factor  of $N_f=2$ explicitly to
facilitate comparison with the dynamical fermion  configurations that
have $N_f=2$.

\begin{figure}[bt!] 
\vspace{11cm} 
\includegraphics{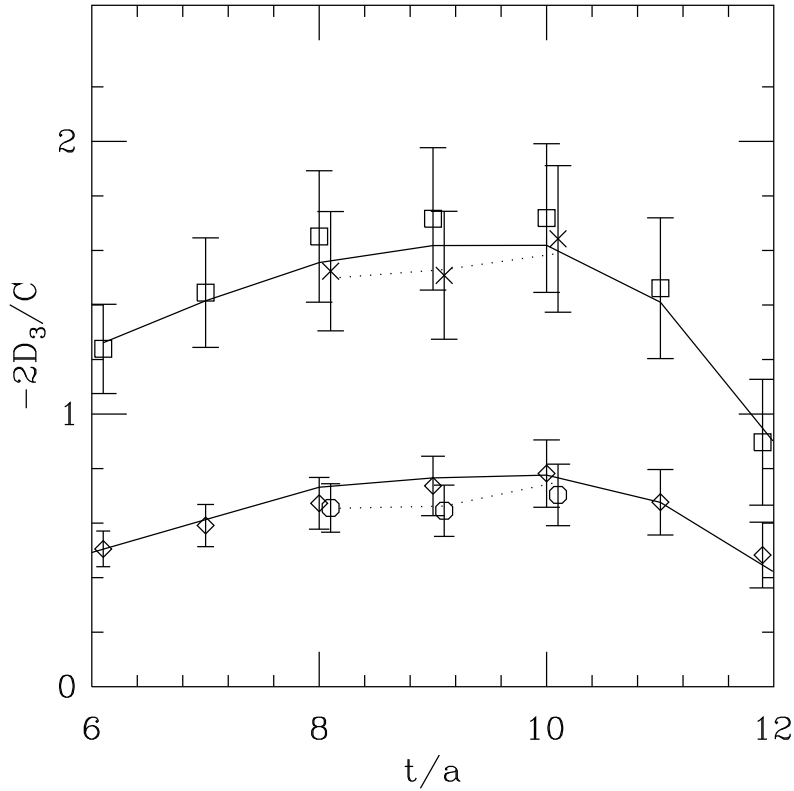}
 \caption{ The disconnected correlation $-N_f D_3/C$ with $N_f=2$ versus
 time $t=t_1+t_2$ in lattice units. The upper data are from dynamical
fermions  with $\kappa_{\rm sea}=0.139$, while the lower data from
quenched  lattices with $\kappa=0.13843$. The curves show the two-state
fits  to these data with $|t_1-t_2|=0$ or  1 as described in the text. The
additional points (crosses  and octagons) have $|t_1-t_2| =2$ or 3  and are
fitted by the dotted curve.
   }
\label{fig.d3r}
\end{figure}

 As discussed previously, the disconnected 3-point correlation
$D_3(t_1,t_2)$  can be fitted to obtain the matrix element $d_{00}$ that
gives us $dM/dm_s$. Because we  only have data on the correlations from local
hadronic operators in this study, we choose to make use of the results
of conventional studies of the 2-point correlators from  both local and
non-local (smeared or fuzzed) operators from larger samples of
configurations~\cite{hugh97,ukqcddf}  to determine the couplings $c_i$ of the
ground state and excited state mesons to our operators.  
 We find that adequate two-state fits can be made to these 2-point
correlations  for $t> 2$. Then keeping the masses and  coefficients
$c_i$ fixed, we can fit all the 3-point data with  $t_1>2$ and $t_2>2$. 
Some typical fits are shown in fig.~\ref{fig.d3r}. The fit results are
shown  in table~\ref{table.disc}: the upper two lines are from quenched 
configurations while the lower three lines are with dynamical fermions.

\begin{table}[hbt]

\caption{Disconnected loop correlations.}

\begin{center}
\begin{tabular}{llllllcc}

 $\kappa$ &  $ M_P$   &  $ dM_P/dm_s$  & $ M_V$ &   $dM_V/dm_s$ & $R_0$ 
    & $n_{\rm gauge}$ & $n_Z$ \\
\hline
 0.14077 &0.529(2) &  1.18(26) & 0.815(5) &   1.6(5) & 2.92(1)&  20 & 
 $16\times 2$ \\
 0.13843 &0.736(2) &  0.86(15) & 0.938(3)  &  1.0(2) & 2.92(1)&  20 &
 $16\times 2$ \\
\hline
 0.1398 & 0.476(14)&  3.0(5)  &  0.706(16) &  2.7(6) & 3.65(4) & 20 &
 $32\times 2$ \\
 0.1395 & 0.558(8) &  3.1(6)  &  0.786(9) &   3.0(7) & 3.44(6) & 20 &
 $32 \times 2$ \\
 0.1390 & 0.707(5) &  1.9(4)  &  0.901(10) &  1.8(3) & 3.05(7) & 24 &
 $20 \times 2$ \\
\hline

\end{tabular}
\label{table.disc}
\end{center}
\end{table}

 The sign of the effect implies that the loop (${\cal T}={\rm Trace}
{\cal M}^{-1}$) is anti-correlated with the pion two-point correlation
$C$ which straddles it in time on a lattice.  This anticorrelation is
large with, for example,  
 \be
 \langle\delta C \ \delta {\cal T}\rangle / (\langle (\delta C)^2\rangle\
\langle (\delta {\cal T})^2\rangle)^{1/2}  \approx -0.5
 \ee  
 at $t=6$ for both the  dynamical fermion and quenched cases.

\begin{figure}[bt!] 
\vspace{11cm} 
\includegraphics{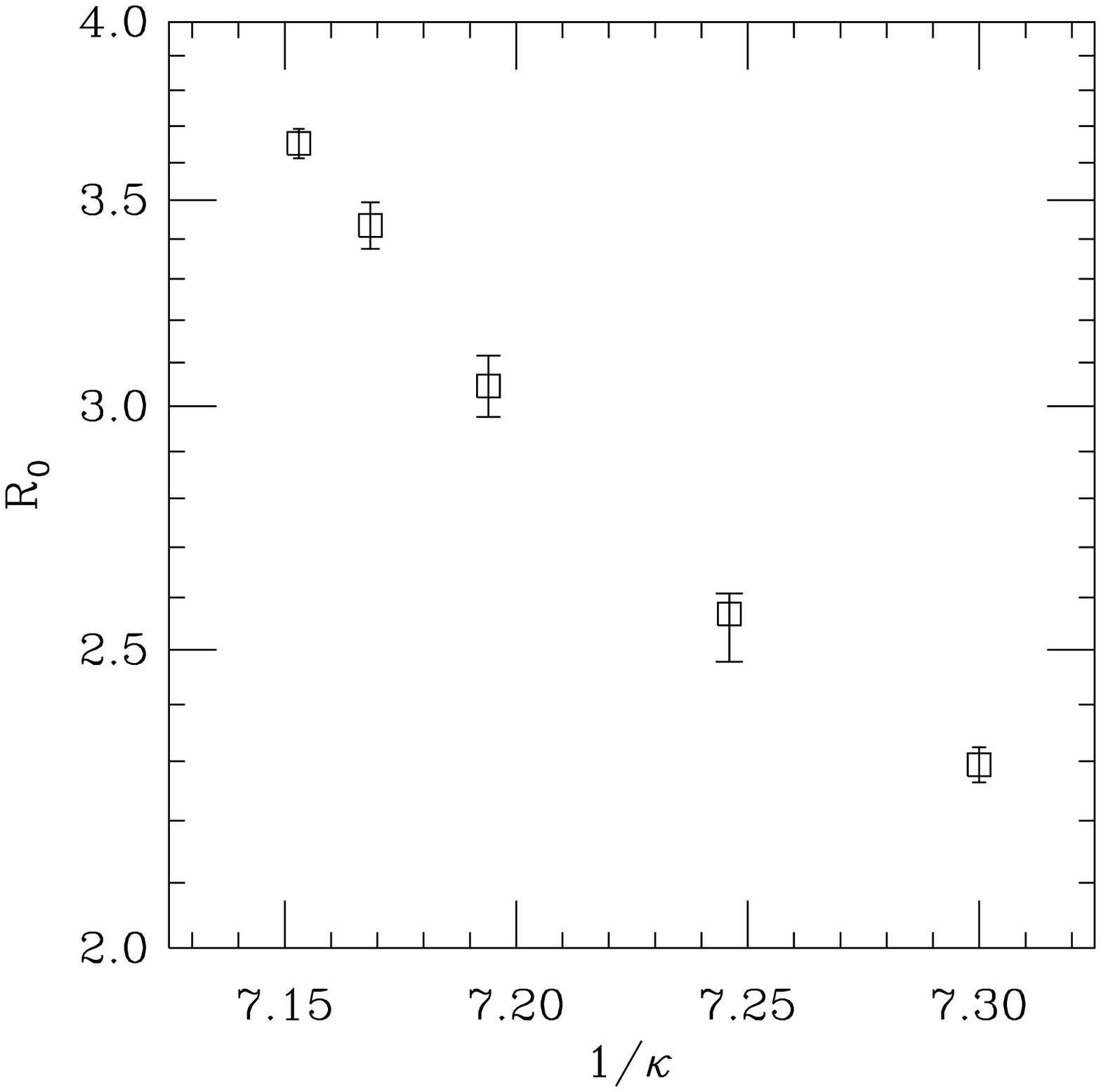}
 \caption{ The  lattice evaluation of  $R_0$  for dynamical
fermion configurations with sea quarks of hopping parameter $\kappa$
from  ref{\protect\cite{ukqcddf}}. We define $m_s \equiv 1/\kappa$.
   }
\label{fig.avsms}
\end{figure}

 This anti-correlation is seen to be very similar for pseudoscalar  and
vector mesons. One qualitative argument for the sign of the  correlation
is that an upward fluctuation  of $C$ corresponds to configurations  in
which quarks propagate easily over large distances whereas an  upward
fluctuation of the loop (${\cal T}$) comes from configurations in which
quarks  do not propagate easily - and so have a bigger amplitude at the
origin. In terms of our identities which relate this disconnected
correlation  to the derivatives $dM/dm_s$, we see that the main effect
comes from the  dependence of the lattice spacing $a$ on $m_s$ at fixed
$\beta$. It is well  known that $a$ decreases as the sea-quark mass
decreases: indeed this is  why the $\beta$ value used in dynamical
simulations is smaller than that used in quenched. The UKQCD
study~\cite{ukqcddf} of the dynamical fermion configurations we are
using finds  $d\,\log a/dm_s \approx -4$ - as shown in 
fig.~\ref{fig.avsms}.  Furthermore the slope appears larger at smaller
sea quark mass - in line  with what we find in table~\ref{table.disc}. 

 We also measure the same disconnected correlation in quenched
configurations.  The results are qualitatively similar to those from
dynamical fermion configurations.  This implies that one can explore the
sea-quark dependence of meson masses  using {\em quenched}
configurations. This appears a striking advance - one can get at
essential information  concerning sea-quarks without the heavy
computational overhead of dynamical fermion simulations. However, it is
widely appreciated that  most lattice observables  are insensitive to
the presence of sea quarks  if the lattice spacing and $M_P/M_V$ ratio
are lined up. Hence, once one has expressed the quantity of interest as 
a vacuum expectation value, it may be evaluated using quenched
configurations. We now explore this in a little more detail.

 For the heavier  quenched ($\kappa=0.13843$) and dynamical
($\kappa=0.1390$) cases,  the lattice spacings (taken from $R_0 \approx
3$) and the $M_P/M_V$ ratio (at 0.78) are very similar. Thus we may 
directly compare the $dM/dm_s$ values obtained.  From
table~\ref{table.disc},  we see  that $dM/dm_s$ has   significantly
smaller values (by two standard deviations)  in quenched  than in
dynamical fermion configurations. This conclusion is reinforced by the
presentation  of the fits to these data shown in fig.~\ref{fig.d3r}. 
These data suggest that this  observable is indeed capable of
distinguishing between configurations  with different sea quark
structure.  One note of caution is that since the lattice spacing is
rather coarse,  the different finite lattice spacing effects in quenched
 and dynamical simulations may be partly responsible for  this observed
difference.  
 
This ability to distinguish quantitatively between quenched and
dynamical  gauge configurations is important - in most cases previously
studied,  no such discrimination was detectable. That the observable
currently under study allows this discrimination  is not entirely
unexpected since the  dynamical fermion configurations are weighted by
${\rm det}({\cal M})$ which is closely  related to ${\rm Trace} {\cal
M}^{-1}$ which is a component of the disconnected correlator.

 For dynamical fermions  it is possible to evaluate by conventional 
methods the hadronic mass differences  as the sea quark mass is varied
and so obtain an estimate of the derivative which can be compared with
our results.
 For the pseudoscalar meson, finite difference
determinations~\cite{ukqcddf}  at fixed valence quark mass 
 of $\kappa_v=0.1390$ give  $dM_P/dm_s=3.1(3)$ and 
 of $\kappa_v=0.1395$ give  $dM_P/dm_s=4.3(4)$.
 Both of these finite difference estimates are somewhat larger than the
derivatives  determined  above. The situation is the same for the 
vector meson mass derivatives where the finite difference determinations
give 3.4(5) and 4.0(1.2) at a fixed valence mass of 0.139 and 0.1395,
respectively.

The two different approaches to determining these sea-quark mass
derivatives used different gauge ensembles  (propagators from the origin
 from about 100 gauge configurations  for ref\cite{ukqcddf},  compared
to propagators from all sites on about 20 gauge configurations here) and
the  differences are only  at the two standard deviation level. At
present the  quoted statistical errors from the derivative method we use
are comparable to those from  finite differences of masses. Using the
full set of gauge configurations  available,  our derivative method
would give the  more accurate determination of the sea quark dependence
of the meson masses.

\section{Discussion}

 There are several issues of interest in determining the dependence  of
hadron masses on the quark masses. Here we are not concerned with the 
problem of defining precisely the quark masses. Rather we discuss the 
dependence of the meson masses as the sea quark mass is reduced to look 
for explicit signs of different physics as the  quark loops become  more
important in the vacuum. One of the complicating features in the lattice
approach  is that changing the sea-quark mass parameter has several
consequences -  among them that the lattice spacing is changed.

As an illustration,  since the lattice spacing $a$ depends on the sea
quark parameter $m_s$,  let us consider  the dimensionless ratio
$M_V/M_P$. Then
 \be
 {d \over dm_s} \ln ({M_V \over M_P}) = {1 \over M_V} {dM_V \over dm_s}
-{1 \over M_P} {dM_P \over dm_s}
 \ee 
 can be evaluated. Since we find $dM_P/dm_s \approx dM_V/dm_s$, this 
gives a negative result which implies that the $M_V/M_P$  ratio
increases as the sea quark mass is decreased. This change  in sea-quark
mass parameter is at a constant $\kappa_v$, however,  which is not 
necessarily what is required.

To clarify this discussion,  it  must be remembered   that the
bare parameters $(\beta, \kappa_v, \kappa_s)$ which occur in the
lattice  formalism are not simply related to the more physical
parameters $a$ and the  sea and valence quark masses which we denote
here as $q_v$ and $q_s$. One example  of this intricate relationship is
that as $\kappa_s$ is increased (i.e.  towards $\kappa_c$ so that the
sea quarks are lighter) then $a$ becomes smaller (this can be seen from
the  observation that $\beta$ needs to be reduced for dynamical fermions
 to keep $a$ approximately the same). Furthermore, this change  of
$\kappa_s$ is also likely to  result in a different value of $\kappa_c$ 
(here defined as the value which gives a massless pion on varying
$\kappa_v$ at that sea quark mass)  and hence the relationship of
$\kappa_v$ with $q_v$ will be modified too.

 Thus one needs to set up a prescription to determine appropriate values
 of the lattice parameters. One proposal is to identify physical
quantities which  should not depend on all of the physical parameters.
Thus we can choose to use  $r_0$ (defined via the static potential at
moderate  separations) to determine the lattice spacing $a$, assuming it
to  be independent of the quark masses. This would not have been true if
the  string tension were to have been used to set the scale since the
string breaking  at large separation will be strongly affected by the
sea quark mass.

  For the quark mass dependence, we are considering a world where the 
valence quark mass $q_v$ can be varied independently of the sea quark
mass $q_s$.  This is not so far from experiment if one regards the $u,d$
quarks as sea quarks  and the strange quark as a valence quark whose
contribution to the sea is relatively small.

 To isolate the quark mass dependence, we choose to make use of a very
conspicuous  experimental fact: the vector mesons are `magically mixed'
with the $\phi$  meson being almost pure $\bar{s} s$ while the $\rho$
and $\omega$ are almost  degenerate and composed of $u,d$ quarks.
Furthermore the $\phi$ has  much reduced decay matrix elements  to final
states containing only $u,d$ quarks. This is the OZI rule: disconnected
quark  diagrams are suppressed. All of this phenomenology suggests that
the vector  meson nonet is well described by the naive quark model: it
does not  contain significant sea quark contributions to the masses.
Thus we choose to  define the sea quark mass $q_s$ such that the vector
meson masses are independent of it. For other mesons, especially the
pseudoscalar mesons,  we do expect some dependence of the masses
explicitly on the sea-quark mass  $q_s$ and we  shall try to estimate
it.

 One way to proceed is to remove the explicit $a$-dependence of the 
lattice masses by forming the product with $R_0$. Then $R_0 M_P$ will 
be equal to the continuum product $r_0 m_P$ up to lattice artefact 
corrections which are of order  $a$ for the Wilson fermion
discretisation but  the clover-improvement scheme we use should reduce
these lattice artefact  corrections to being dominantly of order  $a^2$.
For  ease of notation we define $P_s=d(R_0 M_P)/dm_s$ etc. Here we 
assume, as discussed above, that $R_0$ is independent of $m_v$ and that 
it does depend on $m_s$ through the dependence  of $a$ on $m_s$. This
sea quark mass dependence of $R_0$ can  be extracted by explicitly
evaluating $R_0$ at a range of $m_s$ values~\cite{ukqcddf}, as
illustrated in fig.~\ref{fig.avsms},  giving
 \be
   {1 \over R_0} { d R_0 \over d m_s} = -4.7(1.8) \ ; -4.0(2.0)
 \ee
 where differences are taken from $\kappa_s$ of 0.1390 to 0.1395 and then 
0.1395 to 0.1398 respectively. These values can then be used to obtain
 \be
 {1 \over R_0 M_P} P_s=
 {1 \over R_0 M_P} { d (R_0 M_P) \over d m_s}=
 {1 \over R_0} { d R_0 \over d m_s} 
+{1 \over M_P} { d M_P \over d m_s} 
 \ee
 where a substantial cancellation occurs between the latter two terms.
Thus we find  that the resulting errors are sufficiently large that even
the sign of $P_s $ is not well determined. However, the sign of $P_s$
does not necessarily have any direct physical meaning as we now discuss.

\begin{figure}[htb]
\begin{center}
\epsfig{figure=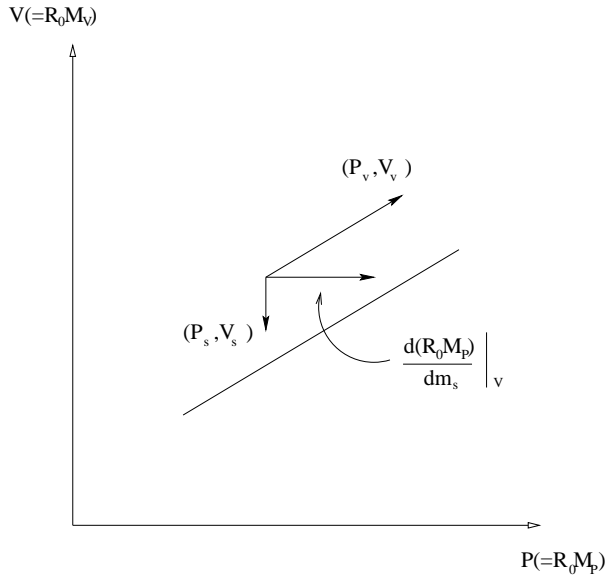,height=3.0in}
\end{center}
 \caption{An illustration of the bare valence ($v$) and sea ($s$) quark 
mass dependence of pseudoscalar ($P$) and vector ($V$) meson masses in 
units of $R_0$. 
 }
 \label{vec.fig}
\end{figure}

 Assuming one had accurate values, we now discuss how to interpret them.
The  situation is illustrated on a plot of $P \equiv R_0 M_P$ against
$V \equiv R_0 M_V$  in fig.~\ref{vec.fig}.

 As the valence quark mass parameter $m_v$ is varied, a curve is traced 
out. What is of interest, however, is the difference between such curves 
as the sea quark mass parameter $m_s$ is varied. Assuming, as discussed above, 
that the vector mass ($V \equiv R_0 M_V$) is independent of $q_s$, then 
yields the required dependence of the pseudoscalar mass on $m_s$ at fixed
$V$:
 \be
  \left. {d(R_0 M_P) \over dm_s}\right|_V = P_s - { V_s \over V_v} P_v
 \ee
 We claim that this quantity will give the physically relevant part  of
the  sea-quark dependence of meson masses: $d(R_0 M_P)/dq_s=r_0
d(m_P)/dq_s)$. Indeed a presentation in this spirit was already shown in
ref.~\cite{ukqcddf}. There it was concluded  that as the sea-quark mass
is reduced, the meson masses move towards closer agreement  with the
experimental data point ($\eta_s$, $\phi$) with  $d(R_0 M_P)/dm_s |_V >
0$ (here $\eta_s $ is the mass expected for a $s \bar{s}$ pseudoscalar
meson).

 Since the precision we obtain in this preliminary study on the
derivatives  is not superior to that which was  obtained by directly
varying the sea and valence masses~\cite{ukqcddf}, the conclusions of
that work are  not modified. However, for dynamical fermion studies
where only one  sea-quark mass is employed, our methods will enable the
derivatives  with respect to the sea-quark mass to be evaluated.

\section{Conclusions}

   One of the current problems in lattice study of hadron spectra, is to
 evaluate the physical consequences of including sea quark effects in
the vacuum. We have presented lattice techniques to evaluate the
dependence of meson masses  on the valence and sea-quark parameters.
These techniques allow such studies to be made  using gauge
configurations at a single set of lattice parameters. This  is a
significant advance for  dynamical fermion studies which are very
computationally intensive. Moreover it  implies that some estimates of
these sea-quark properties can even be made  using quenched
configurations.

  We have discussed how to extract physically useful information about the 
sea-quark effects from these observables. Our proposal takes account of the 
changes induced in the lattice spacing and in the valence mass definition 
as the sea quark parameter is changed. 

 One rather encouraging feature is that we see evidence for a
significant difference  for the disconnected correlation ratio (our
$D_3/C$) between quenched and dynamical quark configurations. It will be
 of interest to explore this difference at finer lattice spacing to
establish that  it is indeed a continuum effect. 

\section{Acknowledgements}

 We acknowledge the support from PPARC under grants GR/L22744 and
GR/L55056 and from the HPCI grant  GR/K41663.

\appendix

\section{Appendix: Analysis of Z2 noise vector methods.}

\subsection{Introduction}

 We summarise first the salient ideas in the Z2
method~\cite{z2liu,sesam}, before  indicating the special features that
we have made use of.

 The required time-slice loop term  can be expressed  in terms of the
quark propagator ${\cal M}^{-1}$ on a given gauge configuration as
 \be
  T(t)=  \sum_x {\cal M}^{-1}_{aj,aj}(x,t;x,t)
 \ee
 where we explicitly show the colour index $a$ and Dirac index $j$ here.
 Since ${\cal M}$ is $\gamma_5$-hermitian, then $T$ is real on any
time-slice  of any gauge configuration. To evaluate this expression for
all $x$ on a time slice using  point sources  would require  solving the
lattice Dirac equation for $L^3$ sources  of each colour and Dirac
index. Let us instead explore using  distributed sources
$\xi^p(x,t)_{bk}$ where $p$ labels the source.
 
Then  solving the lattice Dirac equation from such a  source 
 \be
   G^p_{aj}(x',t')= {\cal M}^{-1}_{aj,bk}(x',t';x,t) \  \xi^p_{bk}(x,t)
 \ee
  and combining with an appropriate combination involving the same source, 
we have 
 \be
{\cal T}^p(t) =  \sum_x \xi^p_{aj}(x,t)^*  \, G^p_{aj}(x,t) 
 = \xi^p_{aj}(x,t)^* \,{\cal M}^{-1}_{aj,bk}(x,t;x',t') \  \xi^p_{bk}(x',t')
 \ee
 Interpreting the sources $\xi$ as random with specific  properties then
makes this quantity, averaged  over realisations of the random
source,  to be just that required, namely
 \be
  \langle {\cal T}^p (t) \rangle = T(t)
 \ee
 This  allows the possibility of  an unbiased estimate of the required
quantity with a moderate  number of inversions ($n_Z$). We require that
the  random sources $\xi^p_{aj}(x,t)$  with $p=1,\dots n_Z$  are such
that  the only non-zero expectation  values of bilinears are given  by
 \be
   \langle  \xi^p_{aj}(x_1,t_1)^* \  \xi^q_{bk}(x_2,t_2) \rangle = 
 \delta_{pq} \delta_{ab} \delta_{jk} \delta_{x_1,x_2} \delta_{t_1,t_2}
 \ee
 This can be implemented by assigning an independent  random number to
each  site $x$, colour $a$ and Dirac index $j$ for each sample $p$. The
optimum distribution of those random numbers can be chosen to minimise
the variance  of the required observable.

 The variance of this estimator  is minimised~\cite{z2liu} by taking Z2
noise (more  correctly Z2$\times$Z2) , namely each component (for real
and  imaginary parts separately) to be randomly $\pm 1/\sqrt{2}$. Then
 \be
   \sigma_z^2= {1 \over 2} {\rm Real}  \left( 
{\cal M}^{-1}_{ij} {\cal M}^{-1}_{ji} +
{\cal M}^{-1}_{ij} {\cal M}^{-1*}_{ij}
 \right)_{i\ne j}
  \ee
 where only the off-diagonal part of ${\cal M}^{-1}$ contributes and here 
we include  space, time, colour and Dirac indices into $i$.

 The variance can be reduced by using a more selective source, for
example~\cite{sesam}  with specific Dirac components. This involves more
inversions, however, if the  full signal is to be evaluated. Here we
choose, instead, to use a source which  is only on a specific time-plane
$t_0$.  Thus in the above formalism  $\xi^p(x,t)_{bk}$ is to be taken as
zero outside the time-slice $t_0$ of interest. This reduces the variance
by a factor  of approximately 4 at the expense of 24 (in our case) times
as many inversions. This is  not cost-effective for evaluating $T$ but
it does enable us to extract  mesonic two-point correlators as we now
discuss.

\subsection{Meson correlators}

   It is also possible to use Z2 source methods to determine meson
correlators. For illustration, consider the correlator between local
hadron operators  of zero momentum given by the average in the gauge 
configurations: 
  \be
 C(t)= \langle 0| H(t_1) H^{\dag}(t_2) | 0 \rangle
 \ee
 with $t=|t_1-t_2|$ and where
 \be
 H(t) =  \sum_x \bar{\psi}_{aj}(x,t)  \Gamma_{jk} \ \psi_{ak}(x,t)
 \ee 
 creates a meson with quantum number given by the Dirac matrix $\Gamma$, 
where $\Gamma=\gamma_5$ for  pseudoscalar mesons and $\Gamma=\gamma_i$ 
for vector mesons.

 Then, using the $\gamma_5$-hermitian property of the fermion matrix, we need 
to evaluate (suppressing the colour indices)
  \be
    C(t)= \sum_{x_1,x_2} \langle 0|(\Gamma \gamma_5)_{ij} 
                             {\cal M}^{-1}_{ik}(x_1,t_1;x_2,t_2)
      (\Gamma \gamma_5)_{kl} {\cal M}^{-1*}_{jl}(x_1,t_1;x_2,t_2)
   | 0 \rangle
  \ee

 This can be evaluated using Z2 methods where $G^p(t)$ is the propagator
from source $\xi^p$ on time-slice $t$ provided one also has the 
propagator $G^{p\Gamma}(t)$  from source $(\Gamma \gamma_5)\xi^p$ on the
same time-slice. Then the  average over samples $p$ of this source will
give the contribution  to $C(t)$ from one time-slice on one gauge
configuration:
   \be
    C(t) = \langle \sum_{x_2} G_i^p(t_1;x_2,t_2)G_j^{p\Gamma *}(t_1;x_2,t_2)
  (\Gamma \gamma_5)_{ij} \rangle
   \ee
   This method allows us to obtain mesonic correlators from any time
slice  to any other. For the pseudoscalar meson, no additional
inversions are  needed since $\Gamma \gamma_5=1$ in that case. For the
vector meson case,  we use $\Gamma=\gamma_i$ with $i=1,2$ or 3 randomly
chosen for each sample $p$. 

 In principle, one could obtain mesonic correlations using Z2 methods without 
additional inversions - for example by explicitly evaluating the average over
samples $p,\ q$ of 
 \be
\sum_{x_2}G^p_i(t_1;x_2,t_2) \Gamma_{ij} \xi^{q*}_j(x_2,t_2)
   \times
\sum_{x_1}G^q_k(t_2;x_1,t_1) \Gamma_{kl} \xi^{p*}_l(x_1,t_1)
 \ee
 In this case, combinatorial factors make the variance of this 
estimator comparable to the signal so it is an inefficient estimator. 
Using sources at  all $t$ would aggravate this problem considerably.

 As described in the text, we can combine the Z2 estimate of the loop 
at $t_0$ with the Z2 estimate of the mesonic correlator, provided $t_0$
is not  the source point of the mesonic  correlator determination. This
restriction is of no  consequence since we are interested in a loop
roughly midway along the  mesonic correlator. 

\subsection{Propagators from Z2 sources}.

    The techniques used to evaluate the propagator from a given  Z2
source are just those used in a standard inversion from  any source. 
This is achieved by an iterative inversion process (either minimal
residual or BiCGstab  algorithms were used). The special feature is that
the precision needed in this iteration is such that any  biases are at
a level substantially below the statistical noise from the  Z2 method.
We are able to monitor several quantities of interest  (eg. ${\rm Trace}
{\cal M}^{-1}$ on a time slice  and the pion propagator to large $t$)
continuously  during the iterative inversion process. The convergence of
these quantities of interest  during the iterative process is not
monotonic, but we are able to establish  a value of the residual that
guarantees sufficiently small systematic errors  from lack of
convergence.  In practice we need approximately one half of the 
number of iterations used in a conventional inversion. This has 
also been discussed by the SESAM collaboration~\cite{sesam}.

\end{document}